\documentstyle[epsfig]{article}
\def \lb#1{\big<#1\big|}
\def \rb#1{\big|#1\big>}
\begin{document}
\date{}
\title{
 Deep Inelastic Structure Functions in  Bag--Like Models
}

\author{
       Jacek Jasiak \\
Institute of Theoretical Physics, Warsaw University
\\ Ho\.za 69, Warsaw, Poland
}
\maketitle

\begin{abstract}
A new method of calculating deep inelastic structure  functions
for the nucleon
in independent particle models is presented. The method
is applied to the bag--like model
and its predictions are
compared with the parametrisation of Gluck, Reya and Vogt.
\end{abstract}

\section{Introduction}
Deep inelastic scattering provides  worthy information
about the structure of nucleons and nuclei. The papers
of Gluck, Reya and Vogt \cite{Reya90,Reya95} have shown that quite
good
agreement  with the data (and moreover some predictions) may
be obtained assuming
valence--like parton distributions at some small
scale $Q^2=\mu^2$ and evolving them using perturbative
QCD. Nowadays it is not possible to calculate the
initial distributions using directly QCD and we are left
with various, less or more phenomenological, models.
The idea of calculating leading twist parton
distributions using quark models has over twenty years
\cite{Jaffe75} and there were many implementations
of it using the bag model, soliton models, the colour--dielectric
model and others \cite{Thomas2021}.

It is reasonable to assume that  interactions between quarks
may be approximately described by  a mean field confining
them.
I fact, even one of the most prominent quark models, the bag model
\cite{bag} (in the static cavity approximation) may be treated
in such a way.

In the mean field approach quarks move independently and there
arises the so called centre of mass problem, resulting from the simple
observation,
that independently moving particles cannot stand for a composite state
of definite total momentum.
The problem manifests itself in the fact, that quark distributions,
calculated in such a model, spread over all Bj\"orken $x$ from
$-\infty$ to $+\infty$ instead of having support in the region
from $0$ to $1$. There are many methods of curing
this problem, but all of them have some drawbacks.
For example the method using Peierls--Yoccoz or Peierls--Thoules
projection gives badly normalised quark distributions.
In this paper we follow the idea of Szymacha \cite{Szymacha}
that the independent particle model
of a composite particle corresponds to the particle moving in an
auxiliary external potential.

\section{Deeply inelastic nucleon structure functions}
Tensor $W^{\mu\nu}$ describing inclusive photon-nucleon scattering
may be written as
\begin{equation}
  \frac{2p^0}{4\pi}\int d^3 \vec x \int d^3 \vec y \sum_n
  \lb{h}J^{\mu \dagger}(\vec x) \rb{n}\lb{n}J^{\nu}(\vec y)\rb{h}
  e^{-i\vec q( \vec x - \vec y)} 2\pi \delta(p^0+q^0-P^0_n),
  \label{wmunu}
\end{equation}
where $q^0$ and $\vec q$ are the photon energy and momentum,
$\rb{n}$ are normalised to $1$ final states with
energy $P^0_n$,
and $\rb{h}$ is the initial (normalised to $1$) nucleon state with
energy $p^0$.

For $\rb{h}$ being an eigenstate of definite momentum $\vec p$
and spin $s$
\begin{equation}
  \rb{h}=\frac{1}{\sqrt{2p^0V}}\rb{h(p,s)},
\end{equation}
where
$\big<{h(p',s')} \big|{h(p,s)}\big>=(2\pi)^32p^0
\delta_{s's}\delta^3(\vec p'- \vec p)  $
and (formally) $V=\int d^3x=(2\pi)^3\delta^3(0)$,\\
we obtain usual form for $W^{\mu\nu}$
\begin{equation}
  W^{\mu\nu}=
  \frac{1}{4\pi} \sum_n
  \lb{h}J^{\mu \dagger}(0) \rb{n}\lb{n}J^{\nu}(0)\rb{h}
  (2\pi)^4 \delta^4(p+q-P_n).
  \label{wmunup}
\end{equation}
If we approximate the real nucleon by independent particles
(as in the bag model), then $\rb{h}$ cannot be a state of
definite momentum and (\ref{wmunu}) is much more useful
then (\ref{wmunup}).

$W^{\mu\nu}$ may be expressed in terms of scalar functions
$F_1$, $F_2$, $g_1$ and $g_2$:
\begin{eqnarray}
W^{\mu\nu}&=&
-(g^{\mu\nu}-\frac{q^\mu q^\nu}{q^2})F_1 +
(p^\mu-\frac{pq}{q^2}q^\mu)(p^\nu-\frac{pq}{q^2}q^\nu)\frac{F_2}{pq}\nonumber\\
&+&i\frac{1}{m}{\epsilon^{\mu\nu}}_{\rho\sigma} q^\rho
\left[ s^\sigma\frac{m^2}{pq}( g_1+g_2)-(sq)p^\sigma\frac{m^2}{(pq)^2}g_2 \right],
\label{deffg}
\end{eqnarray}
where $m$ is the nucleon mass.
 For specific choices of $\mu$ and $\nu$ we obtain
\begin{eqnarray}
  W^{tt}&=W_{\mu\nu}n^\mu_t n^\nu_t=&\frac{F_2}{2x}-F_1, \label{Wfgtt}\\
  W^{\perp\perp}&=&F_1,
\label{Wfgpp}
\end{eqnarray}
where $Q=\sqrt{-q^2}$, $x=Q^2/{2pq}$,
$n^\mu_t=
\left(p^\mu  - \frac{pq}{q^2}q^\mu  \right)/\sqrt{m^2-\frac{(pq)^2}{q^2}}
$,
and $\perp$, $\perp'$ denote components perpendicular
to $n_t^\mu$ and $q^\mu$.

\section{``Naive'' structure functions in the bag model}
As was already stated, independent particles
cannot
correspond to a state which is a momentum eigenstate.
Following Szymacha {\it et.\ al.}\ \cite{Szymacha}
we will assume that they
describe the nucleon with (yet unknown) wavefunction $\Psi_{CM}$,
bounded by an auxiliary external potential.
Nevertheless
our first step is to calculate structure functions of the
bag model (in the static cavity approximation)  without
any improvements for the centre of mass motion.

We assume that in DIS photon interacts only with a given quark from a nucleon,
without any influence on other quarks, exiting it into the
final state $\rb{n_q}$. Then Eq.\ (\ref{wmunu}) may
be rewritten as (we add the subscript $q$ to indicate
that this $W^{\mu\nu}$ is calculated directly from the quark
wave function)
\begin{equation}
\begin{array}{llr}
  W_q^{\mu\nu}=&
  \frac{2m}{4\pi}\int d^3 \vec x \int d^3 \vec y \sum_{n_q}& \\
  &
  \lb{q}J^{\mu \dagger}(\vec x) \rb{n_q}\lb{n_q}J^{\nu}(\vec y)\rb{q}
  e^{-i\vec q( \vec x - \vec y)} 2\pi \delta(p^0+q^0-P^0_n)=  &\\
  \lefteqn{
  m  \sum_{n_q}
   \delta(p^0+q^0-P^0_n)
  } &&\\
 & \lefteqn{
 \int d^3 \vec x e^{-i (\vec q - \vec P_{n_q})\vec x}
 \bar \psi_q(\vec x)\gamma^\mu u_{n_q}
 \int d^3 \vec y e^{i (\vec q - \vec P_{n_q})\vec y}
 \bar u_{n_q}  \gamma^\nu \psi_q(\vec y),
 } &
  \label{wqa}
\end{array}
\end{equation}
where it was assumed that for higher excitation, final states $\rb{n_q}$
may be approximated by plane waves
$u_{n_q}e^{-iP^0_{n_q}x^0+i\vec P_{n_q}\vec x}$,
 and $\psi_q$ is the initial wave function of the quark $q$.

Since bispinor $u_{n_q}$ is uniquely described by corresponding
momentum $\vec P_{n_q} $ and mass
$M=\sqrt{\left(P^0_{n_q}\right)^2 -\left(\vec P_{n_q}\right)^2}$,
we may replace $\sum_{n_q}u_{n_q} \bar u_{n_q} $ by
$\sum _M \int d^3\vec P$ $ \frac{ \gamma P +M}{(2\pi)^3 2\sqrt{M^2+\vec P^2}}$.
For large $Q^2$ we obtain
\begin{equation}
  W^{\mu\nu}_q=
  \frac{m}{4\pi} \sum_M \int da e^{-ip^z a}
  \int d^3 \vec r \bar \psi_q(\vec r)
  \Gamma^{\mu\nu}_q \psi_q(\vec r + \vec e_z a),
  \label{wq}
\end{equation}
where $p^z=\epsilon_q-mx(1+\frac{M^2}{Q^2} )$ ($\epsilon_q$ is the
initial quark  energy),
$\vec e_z=\frac{\vec p }{|\vec p|}$, and
\begin{eqnarray}
\Gamma^{tt}_q &=& \frac{M^2}{Q^2}(\gamma^0-\gamma^z), \label{Gammatt}\\
\Gamma^{\perp\perp}_q &=& \gamma^0-\gamma^z. \label{Gammapp}
\end{eqnarray}
We have to take into account that mass of the exited quark $M$
is the same as the mass of quark in the ground state, hence there
is only one term, with $M=m_q\approx 0$, in the sum $\sum_M$.

\section{Taking into account the centre of mass motion}
The structure functions calculated in the previous section should
be considered as  structure functions for the nucleon in an artificial
external potential. Photon interacting with such nucleon
may excite it into higher eigenstate in the artificial potential,
or excite its internal degrees of freedom, or do the both.
 Since the nucleon is a composite object, its coupling to the photon
cannot be described by simple matrix $\gamma^\mu$. We assume
this coupling to be of the form
\begin{equation}
  f(q^2, P^2, Pq)\gamma^{\mu},
\end{equation}
where $P^\mu$ is momentum of the final state.\\
In general this coupling may have more complicated spinor structure,
but for large $Q^2$ other choices are effectively equivalent
with ours provided we allow $f(q^2, P^2, Pq)$ to be dependent
on the photon polarisation, i.e.\ we take
$f(q^2, P^2, Pq)=f^{A}(q^2, P^2, Pq)$, where $A\equiv \mu\nu=tt$,
$\perp\perp$. 
For large $Q^2$, due to the energy conservation, we obtain
$Pq=\frac{M^2-Q^2}{2}$, so in fact our formfactor depends
only on the polarisation, $Q^2$ and $M^2=P^2$. For a given
 process (fixed $q^\mu$) $Q^2$ remains unchanged and
we may write $W^{\mu\nu}$ for $\mu\nu=tt,$ $\perp\perp$ in analogous
to (\ref{wq}) form (hereafter $ A\equiv \mu\nu $)
\begin{equation}
  W^{A}=\sum_M |f^A(M^2)|^2
  \frac{m}{4\pi} \int da\, e^{-ip^z a}
  \int d^3 \vec r \bar \Psi_{CM}(\vec r)
  \Gamma^{A} \Psi_{CM}(\vec r + \vec e_z a),
\end{equation}
with
 $p^z=E-mx(1+\frac{M^2}{Q^2} )$,
and $\Gamma^{tt}$, $\Gamma^{\perp\perp}$
   given by Eqs.~(\ref{Gammatt}), (\ref{Gammapp}). ($\Psi_{CM}$
  is the nucleon wave function, and $E$ stands for its
  energy.)

Introducing variable $y=1/(1+\frac{M^2}{Q^2} )$,
which by definition satisfies the condition
$0\le y\le 1$,
the above equations may be rewritten in the form
\begin{equation}
  W^A(x,Q^2)=\int_0^1\frac{dy}{y}  W^A_{eff}(y,Q^2)W^A_{CM}(\frac{x}{y},Q^2),
\label{splot}
\end{equation}
where
\begin{eqnarray}
  W^{tt}_{eff}(y)&=& |f^{tt}(y)|^2\frac{M^2(y)}{Q^2}{\cal M}(y),\\
  W^{\perp\perp}_{eff}(y)&=&|f^{\perp\perp}(y)|^2{\cal M}(y),
\end{eqnarray}
(${\cal M}(y)$ is a measure coming from replacing $\sum_M$
by $\int \frac{dy}{y}$, $M(y)=Q\sqrt{\frac{1}{y}-1}$)\\
and
\begin{equation}
 W^A_{CM}(\frac{x}{y},Q^2)=
  \frac{m}{4\pi} \int da\, e^{-i(E-m\frac{x}{y})a}
  \int d^3 \vec r \bar \Psi_{CM}(\vec r)
  \Gamma^{A}_{CM} \Psi_{CM}(\vec r + \vec e_z a),
  \label{WACM}
\end{equation}
with (note difference with (\ref{Gammatt}))
\begin{equation}
  \Gamma^{tt}_{CM}= \Gamma^{\perp\perp}_{CM} = \gamma^0-\gamma^z.
  \label{GammaCM}
\end{equation}

For a constant $\Psi_{CM}$ (corresponding to not moving, unbounded
nucleon), we simply obtain
\begin{equation}
 W^A_{CM}(\frac{x}{y},Q^2)=
  \frac{m}{4\pi}2\pi\delta(E-m\frac{x}{y}).
\end{equation}
In that case $E=m$
and hence from (\ref{splot})
\begin{equation}
  W^{A}=\frac{1}{2} W^A_{eff}.
  \label{WAswobodne}
\end{equation}
Comparing the above expression with (\ref{Wfgtt}), (\ref{Wfgpp}) we obtain
\begin{eqnarray}
  W_{eff}^{tt}&=&\frac{F_2}{x}-2 F_1, \label{Wefftto}\\
  W_{eff}^{\perp\perp}&=&2 F_1, \label{Weffppo}
\end{eqnarray}
This means that to obtain predictions for structure functions in
our model, we have to find $W_{eff}^A$. Since for massless quarks
$W_{q}^{tt}=0$  (the Callan---Gross relation), we are interested
only in $W_{eff}^{\perp\perp}$, which is directly related to the quark
distribution function.
To obtain $W_{eff}^{\perp\perp}$ we have to perform three steps:\\
 a) calculate the naive structure function $W_q^{\perp\perp}$,
 using given quark wavefunction;\\
 b) calculate $W_{CM}^{\perp\perp}$, using the centre of mass
 wavefunction $\Psi_{CM}$
 (function of the nucleon in the auxiliary
 potential);\\
 c) solve the equation Eq.~(\ref{splot}) for $W_{eff}^{\perp\perp}$ using
 found $W_{CM}^{\perp\perp}$ and $W_{q}^{\perp\perp}$ and
 taking $W^{\perp\perp}=W_{q}^{\perp\perp}$.

Of course as far as we don't know the centre of mass
wavefunction $\Psi_{CM}$, this procedure cannot be performed.
We will learn from the next section, how  to find
$W_{CM}^{\perp\perp}$ without calculating explicit form of $\Psi_{CM}$.

\section{Calculating $W^{\perp\perp}_{CM}$}
To obtain $W^{\perp\perp}_{CM}$ from Eq.~(\ref{WACM}), we have to
calculate $ \int d^3 \vec r \bar \Psi_{CM}(\vec r)
  \Gamma^{\perp\perp}_{CM} \Psi_{CM}(\vec r + \vec e_z a)$.
Since $\Gamma^{\perp\perp}_{CM}=\gamma^0- \gamma^z$, we may write
\begin{equation}
\begin{array}{rcl}
 \int d^3 \vec r \bar \Psi_{CM}(\vec r)
  \Gamma^{A}_{CM} \Psi_{CM}(\vec r + \vec e_z a)&=& \\
  \int d^3 \vec r \Psi_{CM}(\vec r)^\dagger
  \Psi_{CM}(\vec r + \vec e_z a)&-&
  \int d^3 \vec r \Psi_{CM}(\vec r)^\dagger
  \alpha^z \Psi_{CM}(\vec r + \vec e_z a).
\end{array}
\end{equation}
The term  $\int d^3 \vec r \Psi_{CM}(\vec r)^\dagger
  \Psi_{CM}(\vec r + \vec e_z a)$ should correspond
  to overlapping function between translated  states of the
  nucleon treated as a composite object. In
  our model the nucleon consist of three quarks with
  wavefunctions $\psi_{q_i}$, so
\begin{equation}
\begin{array}{l}
 \int d^3 \vec r \Psi_{CM}(\vec r)^\dagger
  \Psi_{CM}(\vec r + \vec e_z a)=
 \left(\int d^3 \vec r \psi_{q_1}(\vec r)^\dagger
  \psi_{q_1}(\vec r + \vec e_z a)\right) \cdot
  \\
\left(\int d^3 \vec r \psi_{q_2}(\vec r)^\dagger
  \psi_{q_2}(\vec r + \vec e_z a)\right)\cdot
\lefteqn{\textstyle
\left(\int d^3 \vec r \psi_{q_3}(\vec r)^\dagger
  \psi_{q_3}(\vec r + \vec e_z a)\right) .
}
\end{array}
\label{PsiCMtrans}
\end{equation}
To obtain analogous expression for
$
  \int d^3 \vec r \Psi_{CM}(\vec r)^\dagger
  \alpha^z \Psi_{CM}(\vec r + \vec e_z a)
$
let us remind that for a Dirac particle, mean value of the matrix
$\vec \alpha$ may be interpreted as a mean velocity. It is thus
natural  to take
\begin{equation}
  \vec\alpha_{CM}=\frac{1}{E}
  (\epsilon_{q_1}\vec \alpha_{q_1}+\epsilon_{q_2}\vec \alpha_{q_2}+
  \epsilon_{q_2}\vec \alpha_{q_3}),
 \label{alfaCM}
\end{equation}
where $\epsilon_{q_i}$ is the energy of the quark $q_i$,
and $E$ is the total energy of the system.\\
For identical quark wavefunctions we obtain
\begin{equation}
\begin{array}{c}
 \int d^3 \vec r \Psi_{CM}(\vec r)^\dagger
  \alpha^z\Psi_{CM}(\vec r + \vec e_z a)=\hfill \\
\frac{3\epsilon_q}{E}
\hfill \left(\int d^3 \vec r \psi_{q_1}(\vec r)^\dagger
  \alpha^z\psi_{q_1}(\vec r + \vec e_z a)\right)
\left(\int d^3 \vec r \psi_{q_2}(\vec r)^\dagger
  \psi_{q_1}(\vec r + \vec e_z a)\right)^2  .
\end{array}
\label{PsiCMalpha}
\end{equation}
Equations (\ref{PsiCMtrans}) and (\ref{PsiCMalpha}) mean
that we can calculate $W^{\perp\perp}_{CM}$ just
using $\psi_q$, without introducing $\Psi_{CM}$.

\begin{figure}[htb]
\epsfig{file=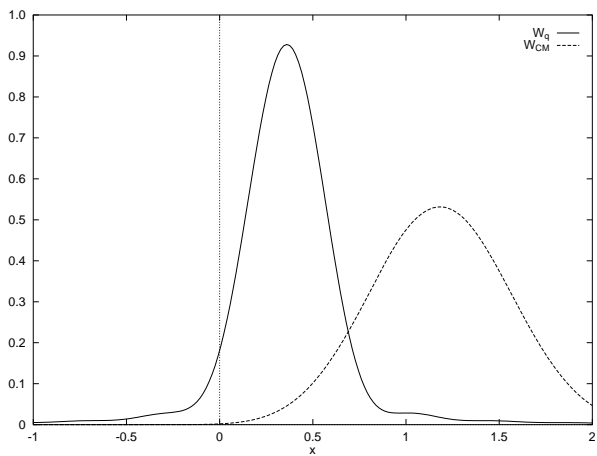,height=8.5cm,angle=-90}
\caption{Functions $W_q \equiv W_q^{\perp\perp}$ and
$W_{CM} \equiv W_{CM}^{\perp\perp}$
in our bag model; $\epsilon_q=2.04\frac{1}{R} $, $E=8.17\frac{1}{R} $, $m=7.62\frac{1}{R}$,
where $R$ is the bag radius.}
\label{rysWqWCM}
\end{figure}

Functions $W_q^{\perp\perp}(x)$ and $W_{CM}^{\perp\perp}(x)$
calculated using the bag model wavefunctions are shown
in Fig.~\ref{rysWqWCM}.

\section{Consistency of the model---sum rules}
Taking Fourier transform of Eq.~(\ref{splot}) we obtain
\begin{equation}
T^{\perp\perp}_q(a)=
\int T^{\perp\perp}_{CM}(ax) W^{\perp\perp}_{eff}(x)\, dx,
\label{splotT}
\end{equation}
where
\begin{eqnarray}
  T_{q}^{\perp\perp}(a)&=&
 e^{i\epsilon_q a}\int d^3 \vec r \psi_{q}(\vec r)^\dagger
  \gamma^0\Gamma^{\perp\perp}_{q}\psi_{q}(\vec r + \vec e_z a), \\
  T_{CM}^{\perp\perp}(a)&=&
 e^{iE a}\int d^3 \vec r \Psi_{CM}(\vec r)^\dagger
  \gamma^0\Gamma^{\perp\perp}_{CM}\Psi_{CM}(\vec r + \vec e_z a).
\end{eqnarray}
For wavefunctions  normalised to $1$ we obtain
$T^{\perp\perp}_q(0)=1=T^{\perp\perp}_{CM}(0)$, and
hence from (\ref{splotT})
\begin{equation}
  \int W^{\perp\perp}_{eff}(x) dx=
  \frac{T^{\perp\perp}_q(0)}{T^{\perp\perp}_{CM}(0)}=1.
\end{equation}
Since $W^{\perp\perp}_{eff}$ has an interpretation as
the quark distribution function, the above equation stands for
obvious normalisation
condition (or from the experimental point of view
Bj\" orken \cite{Bjorken} or Adler \cite{Adler} sum rule).
It should be stressed that  other methods
of obtaining the structure functions from the bag model
have some problems with fulfilling this conditon \cite{Thomas}.

Differentiating
(\ref{splotT}) $n$ times and putting $a=0$ we obtain
\begin{equation}
  \int W^{\perp\perp}_{eff}(x) x^{n-1} dx=
  \frac{{T^{\perp\perp}_q}^{(n-1)}(0)}{{T^{\perp\perp}_{CM}}^{(n-1)}(0)},
\label{momenty}
\end{equation}
where
$
{T^{\perp\perp}_{\dots}}^{(n-1)} =
\left(\frac{d}{da}\right)^{n-1} {T^{\perp\perp}_{\dots}}.
$
The above expression may be rewritten in the form
\begin{equation}
  M_{eff}^{\perp\perp}(n)=\frac{M_{q}^{\perp\perp}(n)}{M_{CM}^{\perp\perp}(n)},
\end{equation}
where
$M_{\dots}^{\perp\perp}(n)=\int W_{\dots}^{\perp\perp}(x) x^{n-1}dx$
(Mellin moments).
It is well known that in the bag model without any
improvements for the centre of mass motion one gets
$2 M_q^{\perp\perp}(2)=\frac{1}{3}$.
This remains true in any independent particle model
with the ``volume'' energy with energy-momentum
tensor of the form:
$$
{\cal T}^{\mu\nu}_V=b(r)g^{\mu\nu},
$$
where $b(r)$ is chosen so to guarantee conservation of the total
(quarks + ``volume'') energy-momentum tensor 
\begin{equation}
  \partial_\nu \left(3\,{\cal T}^{\mu\nu}_q+{\cal T}^{\mu\nu}_V \right)=0.
\label{demuTmu}
\end{equation}
On the other hand energy-momentum tensor ${\cal T}^{\mu\nu}_{CM}$
of our nucleon moving
in the auxiliary potential cannot be conserved, and so
corresponding $2 M_{CM}^{\perp\perp}(2)\ne 1$. But if we modify
the condition (\ref{demuTmu}) by taking into account CM motion
\begin{equation}
  \partial_\nu \left(3\,{\cal T}^{\mu\nu}_q+{\cal T}^{\mu\nu}_V \right)=
  \partial_\nu \,{\cal T}^{\mu\nu}_{CM},
\label{demuTmuCM}
\end{equation}
then we obtain
\begin{equation}
  M_{eff}^{\perp\perp}(2)=
  \frac{M_{q}^{\perp\perp}(2)}{M_{CM}^{\perp\perp}(2)}=\frac{1}{3},
\end{equation}
although in such case $2 M_{q}^{\perp\perp}(2)\ne \frac{1}{3} $.

On the other hand, equation (\ref{demuTmuCM}) together
with (\ref{PsiCMalpha}) allows us to find improved
energy $E$ of our bound nucleon.
For $N$ identical quarks it satisfies the following condition
\begin{equation}
  E=E_0-\frac{N\epsilon_q}{E}\frac{E_0-N\epsilon_q}{N},
\end{equation}
where $E_0$ is the nucleon energy calculated using (\ref{demuTmu}).\\
For the bag model with massless quarks we obtain
\begin{equation}
\begin{array}{ccc}
  \epsilon_q=2.04\frac{1}{R},&
  E_0=4\epsilon_q=8.17\frac{1}{R},& E=7.62\frac{1}{R},
\end{array}
\end{equation}
where $R$ is the bag radius.

\section{Extracting effective quark distributions}
To calculate effective quark distributions, we may use
equation (\ref{splot}) or (\ref{splotT}), but since
$W_{CM}^{\perp\perp}$ and $W_{q}^{\perp\perp}$ are slowly decreasing
functions (see Fig.~\ref{rysWqWCM}),
Eq.~(\ref{splotT}) is much more useful for numerical
calculations. This equation is in fact Fredholm
equation of the first kind,
generally ill-posed --- its solutions may not exist or may
be not unique. The standard method of solving
such problems is to convert them into variational ones
with additional regularization term \cite{numRec}. Following
\cite{illPoss} we introduce a smoothing functional
\begin{equation}
{\cal F}_\lambda[W^{\perp\perp}_{eff}]=
\int \left|T^{\perp\perp}_q(a)-
\int T^{\perp\perp}_{CM}(ax) W^{\perp\perp}_{eff}(x)\, dx\right|^2 da
+\lambda \int \left|W^{\perp\perp}_{eff}(x)\right|^2dx,
\end{equation}
where $\lambda>0$ is a regularization parameter.

Now our goal is to find minimum of the functional
${\cal F}_\lambda$.
Since for physical solution $W^{\perp\perp}_{eff}\ge 0$,
 we impose this condition as
a constraint for  our variational problem.

\begin{figure}[htb]
\epsfig{file=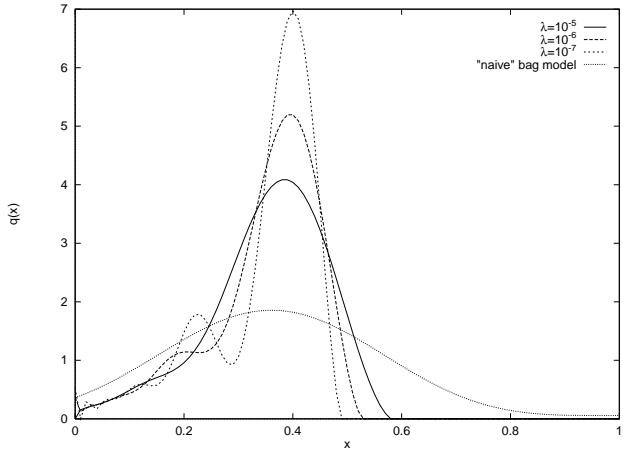,height=8.5cm,angle=-90}   
\caption{Effective quark distributions $q(x)$ in the bag model
found using variational method with different regularization
parameter $\lambda=10^{-5}, 10^{-6}, 10^{-7}$; $\epsilon_q=2.04\frac{1}{R} $,
 $E=7.62\frac{1}{R} $, $m=E$;
For comparison there is also shown ``naive'' quark
distribution $q(x)=2 W^{\perp\perp}_q(x)$
for $\epsilon_q=2.04\frac{1}{R} $, $E=8.17\frac{1}{R} $, $m=7.62\frac{1}{R} $.}
\label{rysWeffa}
\end{figure}

We have solved our problem using procedure {\tt PTIPR}
from \cite{illPoss}.
Results are plotted in Fig.~\ref{rysWeffa}.

\begin{figure}[htb]
\epsfig{file=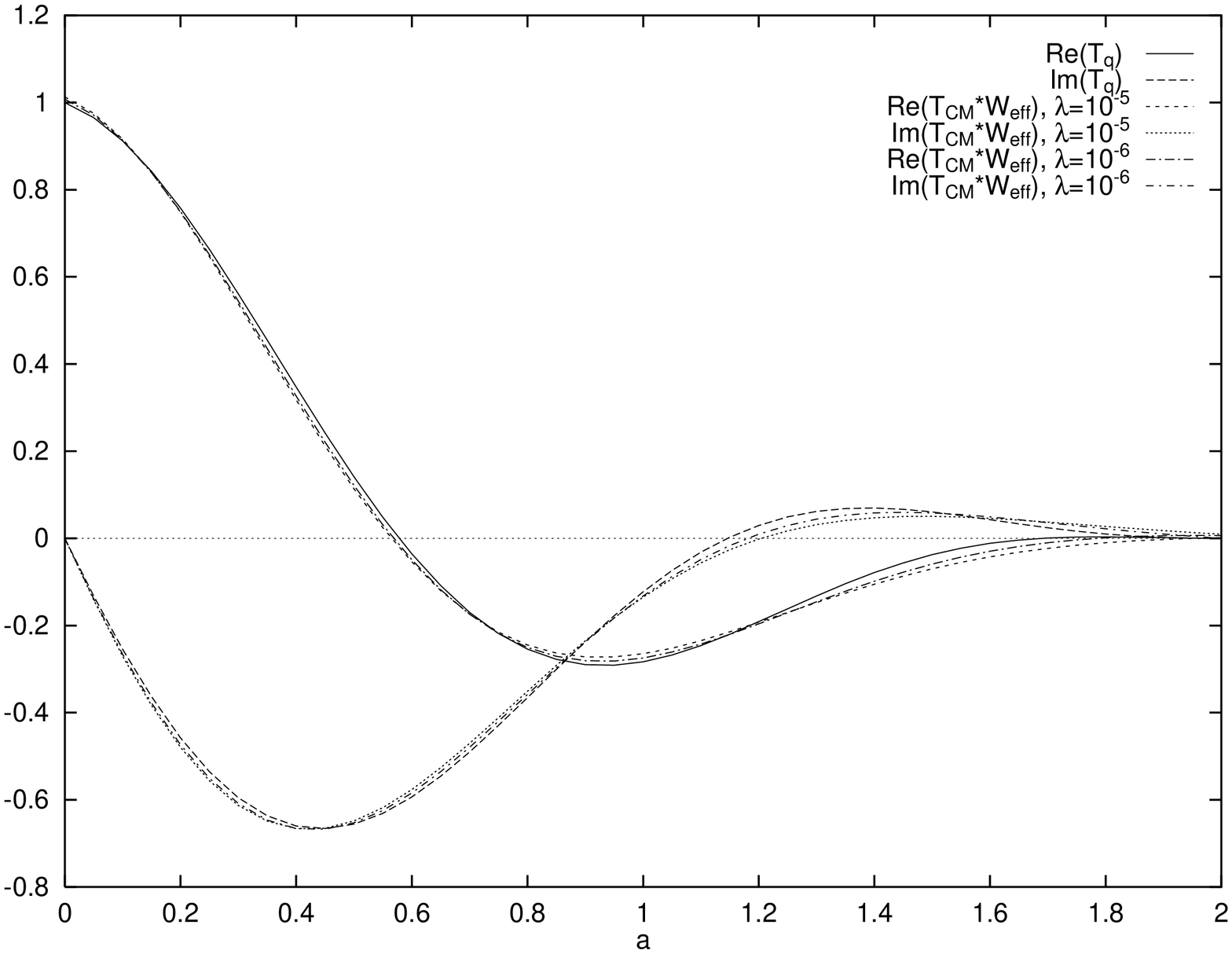,height=8.5cm,angle=-90} 
\caption{Comparison of $T_q$ and $T_{CM} * W_{eff}$
for two solutions for $W_{eff}$ plotted
in Fig.~2 %~{ref{rysWeffa}
}
\label{rysWeffau}
\end{figure}

In Fig.~\ref{rysWeffau} we have plotted  l.h.s.
 of the equation (\ref{splotT}), which can be written
 in the shorthand form as
\begin{equation}
  T_q=T_{CM} * W_{eff}
  \nonumber
\end{equation}
and its r.h.s.\ corresponding to solutions for
$\lambda=10^{-5}$ and $\lambda=10^{-6}$ plotted in Fig.~\ref{rysWeffa}.

\section{QCD evolution}
\begin{figure}[htb]
\epsfig{file=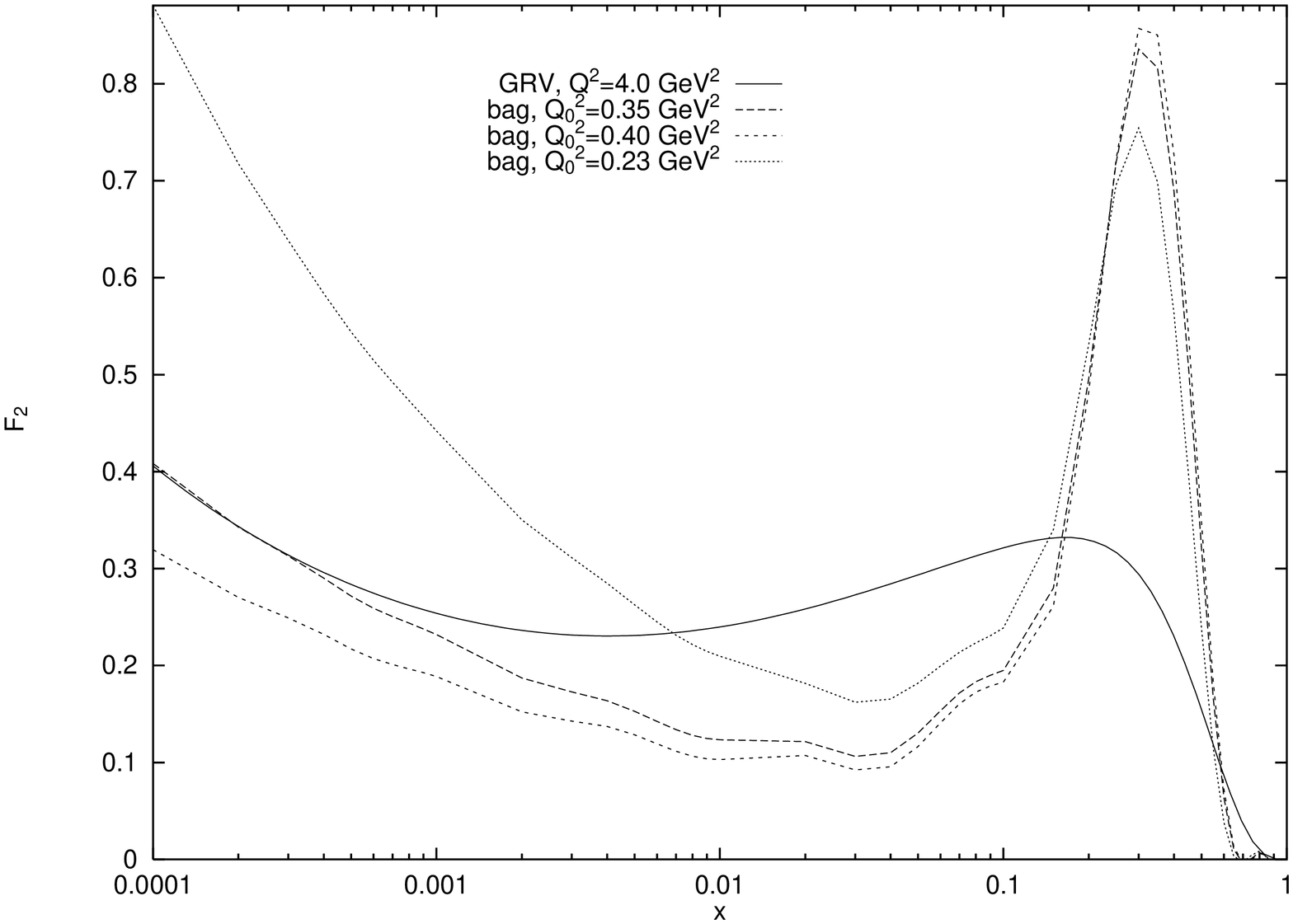,height=8.5cm,angle=-90}
\caption{Results of the LO QCD evolution of
the bag structure functions from $Q^2_0$
to $Q^2=4{\rm GeV}^2$, compared with GRV parametrisation
at $Q^2=4{\rm GeV}^2$. Parameters are:
$\Lambda_{QCD}=200$MeV, $n_f=4$.}
\label{ewolucja}
\end{figure}

The obtained quark distributions in the proton cannot be compared
directly with data. According to the usual procedure
we assume them to be twist two contribution to the
structure functions at small scale $Q^2=Q^2_0$, and then
we
evolve them to a higher scale. Results of the leading order (LO)
evolution
for $F_2$  to $Q^2=4{\rm GeV}^2$ are
plotted in the Fig.\ \ref{ewolucja}.
There also is shown for comparison
 parametrisation of GRV \cite{Reya95}. The chosen initial
scales are $Q^2_0=0.23{\rm GeV}^2$ (the
initial scale of GRV in the LO), and $Q^2_0=0.35{\rm GeV}^2$,
which we found as giving better predictions in the small
$x$ region.

It may be easily seen that evolution of our quark distributions
from $Q_0^2=0.23{\rm GeV}^2$ gives too steep $F_2$ in the
small $x$ region, but it is possible to find $Q^2_0$ giving
much better agreement with the data. On the other
hand, our $F_2$ differs drastically from the data
for $x>10^{-3}$. We attempted to cure this problem
taking into account target mass effects.

\section{Reinterpretation of the results in terms of the
Politzer's variable $\xi$}

Let us start with the remark, that since GRV parametrisation
was obtained by evolving (according to QCD) initial
distributions given at small $Q^2=\mu^2$, we don't have to
perform the evolution ourselves---it is enough to compare
our predictions with that of GRV at some small scale
larger then $\mu^2$.

Described in the previous section procedure of finding quark
distributions
assumes that their entire dependence
(with fixed Bj\"orken $x$) on $Q^2$ comes from QCD.
But due to simple kinetic
target mass effects it cannot be true for smaller $Q^2$.
It was found many years ago \cite{ksi}, that
much better scaling
 may be obtained in terms of the variable $\xi$
 related to the Bj\" orken $x$ by
\begin{equation}
    \xi= \frac{2x}{1+\sqrt{1+\frac{4m_p^2x^2}{Q^2} }},
\label{xi}
\end{equation}
where $m_p$ stands for target (proton) mass.

The connection between usual $F_2(x, Q^2)$ and the
scaling function $F_2^S(\xi, Q^2)$, which for large $Q^2$
coincides with $\frac{F_2(x, Q^2)}{x^2}$, (in that limit
$\xi=x$) is given by \cite{Politzer}:
\begin{eqnarray}
F_2(x,Q^2)&=&\frac{x^2}{(1+4x^2m_p^2/Q^2)^{3/2}}F_2^S(\xi,Q^2)\nonumber\\
          &&+6\frac{m_p^2}{Q^2}\frac{x^3}{(1+4x^2m_p^2/Q^2)^{2}}
          \int_\xi^1d\xi'F_2^S(\xi',Q^2) \label{f2politzer} \\
          &&+12\frac{m_p^4}{Q^4}\frac{x^4}{(1+4x^2m_p^2/Q^2)^{5/2}}
          \int_\xi^1d\xi'\int_{\xi'}^1 d\xi'' F_2^S(\xi'',Q^2),
          \nonumber
\end{eqnarray}
where $\xi$ is the function  of $x$, see Eq.(\ref{xi}).
\begin{figure}[htb]
\epsfig{file=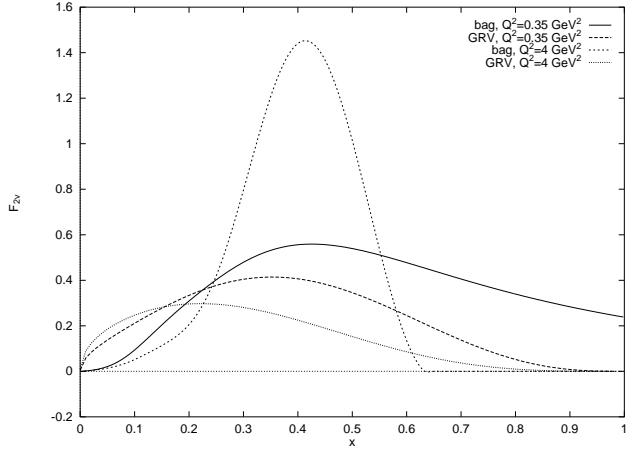,height=8.5cm,angle=-90} 
\caption{
Valence quark part of $F_2(x)$ calculated using Eq.(43) %\ref{f2politzer})
for $Q^2=0.35{\rm GeV}^2$ and $Q^2=4{\rm GeV}^2$ compared
with the parametrisation of GRV.
}
\label{Fxi1}
\end{figure}

$F_2(x,Q^2)$ obtained from the above expression
with $F_2^S(\xi)=\frac{F_2^{eff}(\xi)}{\xi^2}$, where
${F_2^{eff}}(x)=xW^{\perp\perp}_{eff}(x)$  is the effective bag structure
 function obtained in the section 7, compared with GRV parametrisation
for the same $Q^2$, is shown in the Fig.\ \ref{Fxi1}.

\section{Conclusions}
We have proposed here a method of finding hadron structures
functions in  independent particle models.
The method leads to the structure functions
with the correct support and exactly fulfils the
normalisation condition for valence quark distributions.

The structure functions found for the bag model
in the static cavity approximation differs drastically
from the ``naive'' ones, calculated without any
improvement for the centre of mass motion. They
disagree with the data, but  it is possible
to obtain rough agreement in the small $x$ region,
evolving them using perturbative QCD.

Much better agreement with parametrisation
of GRV may be obtained, taking into account
that  structure functions
scale for small $Q^2$ in terms of the Politzer's variable $\xi$.

It should be stressed that our disagreement
with the large $Q^2$ data arises mainly from  quite different
behaviour of our  valence quark distributions at low $Q^2$,
which are finite for $x \to 0$, and the
ones used by GRV, which are singular in that limit
(although their $xq(x)\to 0$). In fact, any independent
particle model gives finite $q_v(x\to 0)$. To
 obtain singular $q(x)$ one has probably to
take into account Regge trajectories for final states
in the deeply inelastic scattering.

\medskip
{\bf Acknowledgements}\\
The author is grateful to Prof.\ M.\ Krawczyk for
many stimulating discussions. He also thanks
 Prof.\ J.\ Kwieci\'nski for
providing him a Fortran code for LO QCD evolution.

\end{document}